\documentclass{PoS}

\usepackage[authoryear,square]{natbib}
\bibpunct{(}{)}{;}{a}{}{,}

\usepackage{amssymb,amsmath}

\title{Identifying the first generation of radio powerful AGN in the Universe with the SKA}

\ShortTitle{Identifying the first powerful AGN}

\author{
\speaker{Jose Afonso}$^{1,2}$, Jordi Casanellas$^3$, Isabella Prandoni$^4$, Matt Jarvis$^{5,6}$, Silvio Lorenzoni$^1$, Manuela Magliocchetti$^7$, Nick Seymour$^8$
\\	
$^1$ Instituto de Astrof\'{i}sica e Ci\^{e}ncias do Espa\c co, Universidade de Lisboa, OAL, Tapada da Ajuda, PT1349-018 Lisboa, Portugal;
$^2$ Departamento de F\'{i}sica, Faculdade de Ci\^{e}ncias, Universidade de Lisboa, Edif\'{i}cio C8, Campo Grande, PT1749-016 Lisbon, Portugal; 
$^3$ Max Planck Institut fur Gravitationsphysik (Albert-Einstein- Institut), D-14476 Potsdam, Germany; 
$^4$INAF - Istituto di Radioastronomia, Bologna, Italy;
$^5$Astrophysics, University of Oxford, Keble Road, Oxford, OX1 3RH, UK;
$^6$Physics Department, University of the Western Cape, Bellville 7535, South Africa; 
$^7$INAF-IAPS, Via Fosso del Cavaliere 100, I-00133 Roma, Italy;
$^8$International Centre for Radio Astronomy Research, Curtin University, Perth, Australia
\\
E-mail: \email{jafonso at oal.ul.pt}
}

\abstract{
One of the most challenging and exciting subjects in modern astrophysics is that of galaxy formation at the epoch of reionisation. The SKA, with its 
revolutionary capabilities in terms of frequency range, resolution and sensitivity, will allow to explore the first Gyr of structure formation in the Universe, in particular, with the detection and study of the earliest manifestations of the AGN phenomenon. The tens of QSOs that are currently known out to the highest redshifts ($z\sim 7$), many of them exhibiting powerful radio emission, imply that super-massive black holes can be grown on a very short timescale and support the existence of very high redshift ($z>7$) radio loud sources - sources that have so far escaped detection. Not only would such detections be paramount to the understanding of the earliest stages of galaxy evolution, they are necessary for the direct study of neutral hydrogen in the Epoch of Reionisation, through observations of the HI 21cm forest against such background sources. 

In order to understand how SKA and SKA1 observations can be optimised to reveal these earliest AGN, we have examined the effect of a hot CMB on the emission of powerful and young radio galaxies. By looking at the SKA1 capabilities, in particular in terms of wavelength coverage and resolution, we determine how the effects of "CMB-muting" of a radio loud source can be observationally minimised and how to identify the best highest-redshift radio candidates. Considering different predictions for the space density of radio loud AGN at such redshifts, we identify the survey characteristics necessary to optimize the detection and identification of the very first generation of radio loud AGN in the Universe.

}

\FullConference{
Advancing Astrophysics with the Square Kilometre Array\\
June 8-13, 2014\\
Giardini Naxos, Sicily, Italy}

\newcommand{\skipthis}[1]{}
\newcommand\nar{New Astron. Rev.}

\newcommand\apj{ApJ}
\newcommand\apjl{ApJ}
\newcommand\mnras{MNRAS}
\newcommand\pasp{PASP}
\newcommand\aap{A\&A}
\newcommand\procspie{SPIE Conf. Ser.}

\begin{document}

\section{Introduction}

The epoch of "the first light" in the Universe, or Epoch of Reionisation (EoR), is one of the most exciting frontiers in current astrophysical knowledge. When (and how) did the first galaxies form? When (and how) did their first stars and supermassive black holes start to shine and how did the first light they produced rapidly ionise the entire Universe? In the quest to understand the earliest phases of galaxy evolution, one requires even more powerful telescopes than available today, capable of reaching faint and rare sources and, even more importantly, to understand how to fine-tune observations for the identification of such "holy-grail" objects. 

Over the last few years, observations have been focusing on two fundamental processes for early galaxy evolution: star-formation and AGN activity. If the radiation from the first stars is often assumed to be the major culprit for the Reionisation of the Universe, recent work \citep{Giallongo2012, Fontanot2012,Robertson2013,Fontanot2014} has raised some doubts over the contribution of accretion to early supermassive black holes (SMBHs). This is particularly relevant to radio observations with the upcoming Square Kilometre Array (SKA), as radio emission from the earliest AGN should be well within its reach. The detection of such very high redshift radio galaxies would be even more exciting for SKA, as it would then be feasible to consider the direct study of neutral hydrogen and its evolution {\it in the Epoch of Reionisation itself}, through observations of the HI 21cm forest against such background sources \citep{Carilli2004,Khatri2010}.

In this chapter we aim to explore how SKA, and in particular SKA phase 1, can be optimised to identify these earliest examples of AGN activity in the Universe. We focus on quantifying the "CMB-muting" of young radio sources - the inverse Compton scattering of synchrotron-emitting electrons off energetic CMB photons, an effect that can be substantial at very high redshifts, and on how SKA radio surveys can be fine-tuned to not only detect but identify the earliest bouts of powerful AGN activity. 
 
\section{The onset of very high redshift AGN activity}

The first sources of light, responsible for the transition from a neutral Universe to a completely ionised one, must have appeared sometime between $z\sim30$ and $z\sim 7$ \citep[e.g.,][]{Barkana2001,Zaroubi2013}.
However, the fundamental observation of sources in the EoR is still only barely manageable: distant powerful quasars, out to $z\sim 7$, showing increasingly strong signatures of neutral hydrogen (HI) absorption in their optical spectra \citep[e.g.,][]{Fan2006,Mortlock2011}; $z\sim 6-9$ star-forming galaxies selected from Lyman-break techniques \citep[e.g.,][]{Bouwens2011,Ono2012,Finkelstein2013,Bowler2014};
the occasional detection (and very limited study) of elusive gamma-ray burst hosts \citep{Tanvir2009,Cucchiara2011}.
Confirmation and detailed investigation of such distant sources is often impossible with current instrumentation, which has prevented expanding our knowledge of the EoR. The very origin of the reionisation mechanism itself is still unclear. Different studies have pointed to star formation as the origin of the photons that ionise the  neutral hydrogen - if enough (proto-)galaxies start forming stars, even if at relatively low levels, and if these stars are hot and energetic enough, a few hundred Myr will be enough to completely ionise the Universe \citep{Robertson2013}. On the other hand, accretion to a SMBH, another process that can produce the necessary ionising radiation, was deemed unlikely, as the time needed to nurture and grow such a "beast" seemed prohibitive, and the apparent break in the luminosity function of powerful AGN at high redshifts made such a significant population of ionising sources unlikely. 

However, there has been an increasing number of detections of high redshift luminous ($L > 10^{44}$\,erg/s) quasars, out to the current record breaker ULASJ1120+0641, at $z=7.1$ \citep{Mortlock2011}. By themselves, these observations show that 10$^8$-10$^9$\,M$_{\odot}$ SMBHs already exist well within the first Gyr of the Universe, implying a very early formation and a surprisingly rapid growth. Since observations are obviously biased towards the most powerful objects, this raises the intriguing question about the ionising contribution of a possibly significant number of slightly less luminous AGN in the first Gyr \citep{Giallongo2012}.

Anchored by these observations, theory has advanced significantly over the last few years (e.g., \citealp{Haiman2004,DiMatteo2008,DiMatteo2012}; see \citealt{Volonteri2010} for a review).
%check Santos, M. G., Silva, M. B., Pritchard, J. R., Cen, R., & Cooray, A. 2011, A&A, 527, A93
The problem is now not if a SMBH can be grown quickly enough (observations show that they can), but how to produce a sufficiently massive (10$^2$ to 10$^5$\,M$_{\odot}$) BH seed a few hundred Myr before. If such BH seeds exist, they can lead to suitable SMBHs at $z>7$, even without a continuous, and difficult to envisage, accretion at the Eddington rate for the entire period of growth. The remnants of the first generation (Pop. III) stars now appear {\it not} to be the such seeds, as sufficiently high stellar masses are less common than previously though \citep{Turk2009,Clark2011,Greif2011,Stacy2012}. 
Suitable seeds are nevertheless possible via direct gas collapse,
quasi-stars produced by very high gas infall rates,
the collapse of star clusters 
or even primordial black holes, formed much before the epoch of galaxy formation 
\citep[e.g.,][]{Carr2003,Bromm2003,Mayer2010,Begelman2010,Devecchi2010,Devecchi2012,Khlopov2010}.

Whatever the process, the end result - the existence of SMBHs well within the first Gyr of the Universe - now seems established, and their detection and study is fundamental to understand the earliest phases of galaxy formation. The SKA, with its revolutionary capabilities, will be fundamental to explore these elusive sources, revealing the first steps of the AGN phenomenon in the Universe and the role of SMBHs to the development of galaxies from the earliest times.

\section{The space density of very high redshift radio powerful AGN}

A crucial ingredient to understand the role of a revolutionary telescope such as SKA for the observation of the first powerful AGN, is an estimate of their expected {\it space density}. Several models, taking different approaches to early AGN evolution, are currently available to produce such estimates. 

A simple physically motivated semianalytic model for the SMBH population out to very high redshifts is presented by \citet{Haiman2004}. Constrained by the optical-IR and X-ray quasar luminosity functions (LFs) at lower redshifts ($z\sim 5$), the Luminosity Function and number counts of bright ($\sim 1$\,mJy) radio sources at high redshift and the counts at the 10\,$\mu$Jy level in deep radio observations, this model predicts the detection of $\sim 60$ AGN per deg$^2$ for $z>6$, $\sim 20$ for $z>8$, and $\sim 10$ for $z>10$, for a radio survey with a detection threshold of 10\,$\mu$Jy. In this model, a discrepancy in the predicted source counts at $\sim 10$\,$\mu$Jy is solved by assuming that SMBHs with masses $M<10^{7}$\,M$_{\odot}$ are either rare or are inefficient at producing radio emission. If that is the case, then the detection rate of very high redshift AGN does not change significantly even if reaching lower observed radio fluxes -- essentially all radio emitting AGN, those with $M>10^{7}$\,M$_{\odot}$, are already detected at 10\,$\mu$Jy in this model. Another point to notice is that a flat spectral index is assumed -- if this is relaxed to $\alpha \sim 0.5$ ($F_\nu \propto \nu^{-\alpha}$) the counts are reduced by a factor of a few, at most, and only for the highest frequencies of $\sim 10$\,GHz. 

Another model appropriate to estimate the abundance of very high redshift AGN is the SKADS Simulated Skies \citep[S3,][]{Wilman2008,Wilman2010}. The model predicts the detection of $\sim 160$ AGN per deg$^2$ for $z>6$, $\sim 100$ for $z>8$, and $\sim 70$  for z>10, for a radio survey complete to 10\,$\mu$Jy. The AGN type separation therein indicates that one third of the detected AGN in the model at $z>6$ are young Gigahertz-Peaked Sources (GPS), while the other two thirds are essentially composed of FRI sources. The AGN mix changes at $z>10$ to a 50/50 ratio between FRIs and GPS sources.

Although the difference between the \citet{Haiman2004} and S3 predictions is significant, both models suggest a very substantial presence of detectable AGN at the highest redshifts. However, in spite of the optimistic predictions, the unavoidable fact is that very high redshift radio AGN have not been easy to detect, even after years of dedicated efforts by several teams. The highest redshift purely radio selected AGN continues to be TN J0924-2201, at a redshift of $z=5.2$ \citep[][]{vanBreugel1999}. More recently, selection based on a radio-to-near-infrared criterion led to the discovery of a $z=4.9$ radio powerful source \citep{Jarvis2009}. The combination of optical quasar selection criteria with a radio detection has also resulted in the identification of a number of radio luminous AGN at high redshifts \citep[$z\sim 6$, e.g.][]{McGreer2006,Zeimann2011}. 

One of the major difficulties is certainly the {\it confirmation} of radio-selected very high redshift candidates, as this relies in lengthy optical/NIR spectroscopy observations with the largest telescopes available. However, given that we currently know a few tens of optical/NIR selected QSOs at $z\sim 6-7$ \citep[e.g.,][]{Fan2006,Willott2007,Willott2009,Jiang2008,Mortlock2011,Venemans2013}, even selection biases start having a rough time explaining the sparseness of very high redshift radio sources. This naturally leads to analysing more deeply the physics of radio emission within the first Gyr after the Big Bang: are there physical processes that will lead to a decreased radio emission that can explain the difficulty in finding the earliest examples of AGN activity? And, if so, can one tune a radio survey in the frequency vs. sensitivity parameter space in order to optimize the probability of finding such early sources?

\section{The physics of continuum radio emission at very high redshift}

The continuum radio emission from AGN is dominated by synchrotron emission produced by highly energetic charged particles powered by the collapse of matter to a SMBH. In the early Universe, this process will happen with two important differences with respect to what happens at more recent epochs: the environment will be denser and the interaction with the cosmic microwave background (CMB) will be more significant, given the higher energy density of its radiation field ($U_{CMB} \propto (1+z)^4$). 

In a recent work, \citet{Ghisellini2014} have considered the interaction between the extended emission of a radio jet and the CMB, concluding that energy losses of emitting electrons by Inverse Compton (IC) to the hot CMB may overcome synchrotron losses for epochs earlier than by $z\sim 3-5$ (an effect we will call here CMB-muting, and which can easily reach factors of 10 at GHz frequencies, depending on the strength of the magnetic field in the emitting region). This is more significant at higher radio frequencies, which suggests that finding powerful AGN at high redshift based on the identification of extended radio structures should be better performed at lower radio frequencies (e.g., observing in the $10-100$\,MHz range instead of $\sim 10$\,GHz).

Given that early galaxy formation happened in much denser environments, the expansion of radio emitting material must also have been more difficult. Hence, environment, age, and significant CMB-muting for very high redshifts, all combine to favour looking for compact radio sources in the highest redshift Universe. Even if the jets are able to expand and form extended lobes, which will likely be severely CMB-muted at the highest redshifts, their compact radio cores may still be detectable as compact radio sources. 

We have recently started looking at the effects of CMB-muting in compact radio sources \citep{Casanellas2015}, namely Compact Steep Spectrum (CSS) sources or Gigahertz Peaked-Spectrum (GPS) sources \citep[e.g.,][]{Falcke2004}. These are radio AGN at their early evolutionary stages, with GPS sources being considered to evolve into CSS sources in a self-similar way at later stages, displaying increasing sizes (up to a few kpc) and radio spectra peaking at lower frequencies \citep[to below $\sim 500$\,MHz, see e.g.,][]{Odea1998}. GPS and CSS sources are thought to be the progenitors of larger classical radio galaxies \citep{Carvalho1985,Readhead1996}, and GPS or CSS-like radio spectra are also found in the inner jets or radio cores of AGNs with extended lobes. 

The main emission mechanism in these radio sources is the synchrotron radiation from the relativistic electrons in their jets, as in classical radio galaxies but with the hot spots located closer to the nucleus. The physical processes governing the energy balance of the electrons in these sources at high redshift are similar to those in the nearby Universe, with the difference that the denser radiation field of the CMB which may play a more important role. 

By modelling the balance between synchrotron emission (including the effects of synchrotron self-absorption, important for sources in their early evolutionary stages, such as GPS and CSS) and CMB-muting in the core of radio-powerful AGN, we find that in contrast with extended sources, compact radio cores such as CSS and GPS sources do not generally suffer significant radiative losses due to inverse Compton with the CMB -- even at the highest redshifts ($z\gtrsim 10$). Only for CSS sources with weaker magnetic fields ($\lesssim 200$\,$\mu$G) do we find significant CMB-muting, resulting in a reduction of the expected flux density at high frequencies ($\gtrsim 1$\,GHz). In this case, the flux density is much less affected at lower frequencies and, with the decrease of the break frequency above which the CMB-muting is most obvious, the radio SED becomes steeper between MHz and GHz frequencies. 

\begin{figure*}[]
 \begin{center}
\includegraphics[scale=0.9]{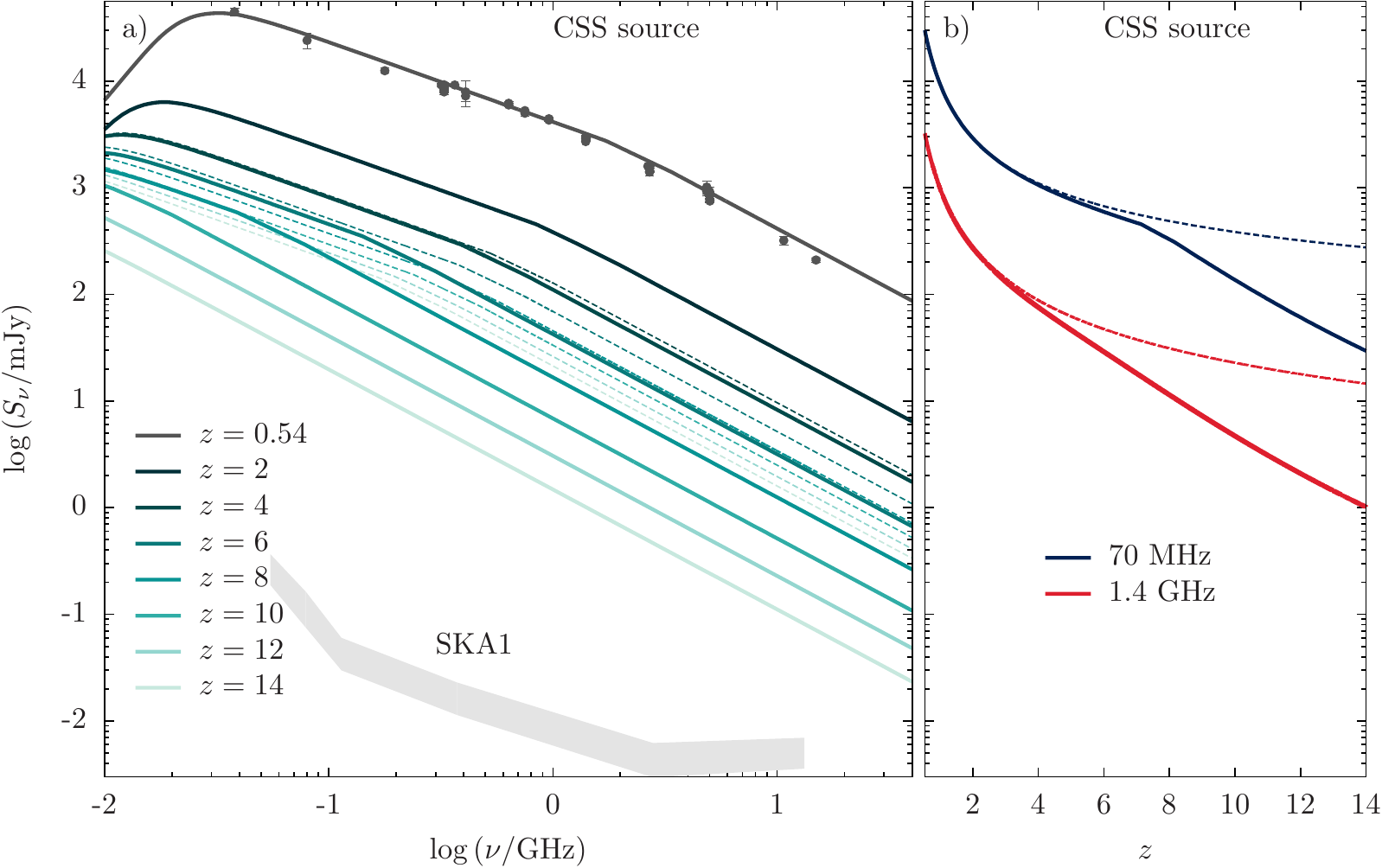}
\end{center}
\caption{a) Observed radio SED of the CSS source, 3C455, if placed at different redshifts. The effect of redshift  (dashed lines) and considering also CMB-muting (solid lines) is shown. A magnetic field of $B=200$\,$\mu$G \citep{Murgia1999} was assumed. The gray region marks the sensitivity limits (5 to 10 $\sigma$) of SKA1 for an integration time of 30 minutes; b) Flux densities at 70 MHz and 1.4 GHz from the same CSS source.}
 \label{fig-fluxCSS}
\end{figure*}

As an illustration of the potential effect of CMB-muting in compact radio sources with weaker magnetic fields, we show in Figure~\ref{fig-fluxCSS} how the observed radio SED of the CSS source 3C455 ($z=0.543$) would change if it was placed at increasing redshifts. The substantial CMB-muting at high redshift steepens the spectrum of the source, strongly reducing the flux density at the higher frequencies. On the other hand, the impact is weaker at low frequencies, as the CMB-muting preferentially reduces the population of the more energetic electrons, the main contributors to the high frequency region of the spectra.

A significant reduction of the flux density at high frequencies would be observed for 3C455 at $z\gtrsim 5$ due to the energy losses by CMB-muting. For example, at $z=8$ the flux density at $\sim 1$\,GHz would decrease by a factor of 3 when CMB-muting is taken into account (see Figure~\ref{fig-fluxCSS}). On the other hand, the reduction of the flux density at 70\,MHz is more moderate at the same redshift.

\begin{figure*}[]
 \begin{center}
\includegraphics[scale=0.9]{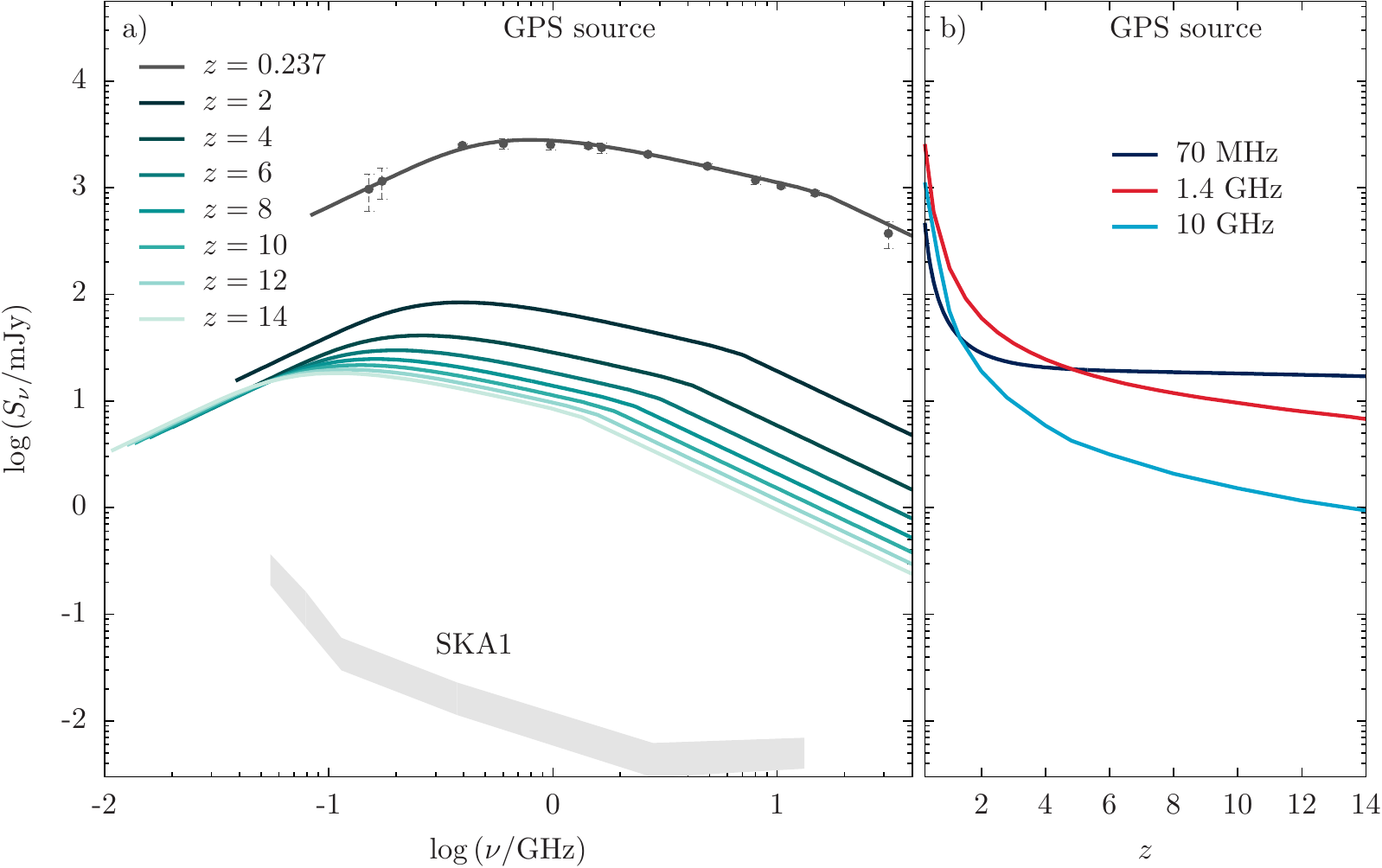}
\end{center}
\caption{a) Observed radio SED of a representative GPS source, GPS 2352+495, at different redshifts, as in Figure~\protect\ref{fig-fluxCSS}. A magnetic field of $B=5$\,mG was considered \citep{Odea1998}. The gray region marks the sensitivity limits (5 to 10 $\sigma$) of SKA1 for an integration time of 30 minutes; b) Flux densities at 70 MHz, 1.4 GHz and 20 GHz from the same GPS source.}
 \label{fig-fluxGPS}
\end{figure*}

We note that such low magnetic fields as those in 3C455 are observed in a very small percentage of the observed CSS population \citep{Murgia1999}. For the vast majority of known CSS and GPS sources, CMB-muting will not have an appreciable effect on their observed radio SEDs. An example is shown in Figure~\ref{fig-fluxGPS}, for the GPS source GPS 2352+495 ($z=0.237$). In contrast to the CSS source, the IC/CMB mechanism does not produce significant radiative losses due to the larger magnetic fields. One should note that the flux density at low frequencies does not decline with redshift as quickly as at high frequencies, pointing to the interest in considering the lower frequency regime ($\sim100\;$MHz) when searching for the highest redshift sources.

Furthermore, it is also noteworthy that the detection of the turnover frequency, due to synchrotron self-absorption, for compact radio sources is particularly relevant for their potential identification from radio observations alone -- fundamental at the highest redshifts. The coverage of the low frequency regime ($\nu\sim 10-500$\,MHz) with a sensitive radio telescope would thus be of particular interest. 

\section{Identifying the highest redshift radio sources with SKA}

In terms of continuum detection of the highest redshift radio powerful AGN, SKA1 should be able to detect thousands of such sources at redshifts $z\sim 6-10$. As seen above, both \citet{Haiman2004} and the S3 models predict that a wide (1000-5000 deg$^2$) survey at $\sim 1$\,GHz reaching a detection level of 10\,$\mu$Jy will reveal as many as $\sim 10^5$ powerful AGN at $z>6$, several tens of thousands at $z>8$ and even $\sim 10^4$ at $z>10$. Even considering the indication of the S3 models that up to half of these sources (the extended FRIs in the sample) may be affected by considerable CMB-muting, that still leaves a few thousand AGN (the young and compact ones) detectable even at $z>10$. Naturally, the huge uncertainties about a possible large decrease of the luminosity function of AGN at the highest redshifts \citep[e.g.,][]{Rigby2011} advise some caution in using these numbers, but, as far as we can tell, even the shallower SKA1 surveys \citep[see][this volume]{Prandoni2014} will reveal a large number of very high redshift powerful AGN. 

As can be seen in Figure~\ref{fig-fluxCSS} and Figure~\ref{fig-fluxGPS}, the exploration of radio observations can provide some indication about the redshift of the source, in particular when considering low-frequency radio observations -- either from the SKA itself, or adding observations from observatories like LOFAR, at even lower frequencies ($\sim 10-100$\,MHz). Nevertheless, the radio SED shape alone will likely not be sufficient to provide more than a way to help selecting potentially interesting sources for follow-up work. In order to confirm that radio sources are indeed at high-redshift (say, $z>6$) multi-wavelength data will be crucial. Deep imaging above and below the Lyman-$\alpha$ line will be required, which consequently means a combination of optical and near-infrared observations as even at $z = 6$  Lyman-$\alpha$ is at $\sim 8500$\AA, and as we push to much higher redshifts then the short wavelength band is pushed well into the near-infrared window. Therefore the key surveys for this science case are LSST \citep[see][this volume]{Bacon2014}, to essentially rule out low-redshift interlopers from the high-redshift radio galaxy sample, and a combination of wide-area, deep near-infrared imaging from {\em Euclid} and deep field near-infrared imaging from either ground-based facilities such as VISTA now \citep[see e.g.][]{McCracken2012, Jarvis2013} but possibly crucially the {\it JWST} which will be able to detect these galaxies individually with relatively short exposures, thus confirming the likelihood of the source being within the epoch of reionisation. Nevertheless, one should not neglect the upcoming availability of powerful near-infrared spectrographs, like {\it MOONS} \citep{Cirasuolo2012} at the VLT, which will be able to confirm a very high redshift nature for many of the brightest candidates.

\section{Conclusions}

The unparalleled capabilities of SKA, in terms of sensitivity but mostly in terms of survey speed and wide frequency coverage, will allow for the detection of young powerful radio galaxies at very high redshifts (even out to $z\gtrsim 8$). The first powerful radio sources, whenever they exist, will be amongst the sources revealed by the early SKA1 surveys, allowing for the study of the first bouts of AGN activity in the Universe. The biggest challenge will be to identify such sources amongst the millions that will be observed. The use of the wide frequency range of SKA, possibly together with other radio observations at lower frequencies, will allow for the identification of robust candidates for such high redshift sources. Still, the combination of SKA surveys with optical and NIR surveys will be necessary for the confirmation of sources well within the epoch of reionisation. Given the availability of such complementary data in the near future, the study of the very first generation of radio loud AGN in the Universe will be well within our reach.

\section*{Acknowledgments}
JA and SL gratefully acknowledge support from the Science and Technology Foundation (FCT, Portugal) through the research grants PTDC/FIS-AST/2194/2012 and PEst-OE/FIS/UI2751/2014, and the fellowship (SL) SFRH/ BPD/89554/2012. JC acknowledges support from the Alexander von Humboldt Foundation. MJ acknowledges support by the South African Square Kilometre Array Project, the South African National Research Foundation.

\end{document}